A neuromorphic vision system for open-world visual intelligence

**Authors**

Jiankai Yin[1 †], Zheng Miao[2,3 †], Lianghao Guo[1 †], Cong Li[4], Shengbo Wang[4], Hongfu Xu[1], Weihao Ma[1], Yuyang Zeng[5], Yubiao Luo[2,3], Yongxiang Li[2,3], Saitao Zhang[2,3], Arokia Nathan[6,7], Luigi Occhipinti[8], Shuo Gao[4*], Zhong Sun[2,3,9*], Xiaoyu Guo[5*]

**Affiliations**

[1] School of Automation Science and Electrical Engineering, Beihang University, Beijing 100191, China
[2] Institute for Artificial Intelligence, Peking University, Beijing 100871, China
[3] School of Integrated Circuits, Peking University, Beijing 100871, China
[4] School of Instrumentation and Optoelectronic Engineering, Beihang University, Beijing 100191, China
[5] Department of Mechanical Engineering, City University of Hong Kong, Hong Kong, China
[6] School of Information Science and Engineering, Shandong University, Qingdao 266237, China
[7] Darwin College, University of Cambridge, CB2 1PZ, Cambridge, UK
[8] School of Mechanical, Electrical and Manufacturing Engineering, Loughborough University, Loughborough LE11 3TU, UK
[9] Beijing Advanced Innovation Center for Integrated Circuits, Beijing 100871, China

[†]These authors contributed equally to this work
[*]Correspondence to shuo_gao@buaa.edu.cn, zhong.sun@pku.edu.cn, xiaoyguo@cityu.edu.hk

**Abstract**

Time-efficient and robust visual intelligence remains a critical challenge in unstructured open-world environments, yet current approaches often rely on computationally intensive neural architectures or task-specific sensors with limited versatility. Inspired by biological vision and information bottleneck theory, we report a neuromorphic vision system that performs task-oriented visual intelligence through an information distillation strategy (named as task traction mechanism) implemented on hardware. The system integrates a polarization-sensitive imager with a resistive random-access memory (RRAM) array to progressively distill task-relevant information via light field selection, region of interest extraction, and target anticipation. The neuromorphic vision system conducts visual tasks within an execution time of 193 $\mu$s. Evaluation across eight challenging open-world scenarios shows accuracy improvements of 25.54%, 37.73%, and 36.10% for object tracking, object segmentation, and trajectory prediction, respectively, together with an average 30.6-fold reduction in latency relative to state-of-the-art solutions.

**Introduction**

Vision-based perception is central to decision-making in autonomous systems [1-5]. However, achieving time-efficient and reliable perception remains challenging in unstructured open-world environments [6-9], as evidenced by the 5.25-fold increase in autonomous vehicle accidents during dusk and dawn, owing to strong luminance variations and glare [10]. Current algorithmic approaches improve reliability by scaling up training datasets [11-15] and adopting complex image enhancement models [16-25], but the resulting computational overhead and inference latency limit their use in real-time application scenarios such as autonomous driving (SNote 1). Recently, task-specific perception based on multidimensional light field information has been widely explored in vision systems [26], including polarimetric [27-29], infrared [30-32], and hyperspectral information [33-35] (SNote 2). While effective towards specific tasks, such as reflection removal, low light thermal sensing, or spectral material separation (Figures 1(a) and (b)), their performance are limited under the diversity of real-world scenes, hence a pathway towards task-adaptive perception is highly expected.

Guided by information bottleneck theory, preserving task-relevant information while compressing the input has been identified as a promising pathway towards enhancing the efficiency of visual information processing in unstructured scenes [36, 37]. At the system level, this can be achieved by selectively distilling light field information according to task demands (SNote 3). A biological archetype for this information distillation is found in the visual pathway of the dragonfly: first, the dragonfly selectively utilizes light field intensity and polarization according to environmental demands, in which polarization provides critical information on surface geometry, material properties, and scattering [38, 39]; second, its visual pathway identifies task-relevant regions through the spatiotemporal filtering of large monopolar cells (LMCs) [40-42]; finally, they predict the short-term trajectories of moving targets through the small target motion detector (STMD) [43, 44] (SNote 4). Through this pathway, the dragonfly preserves information relevant to the target while compressing other visual inputs, which supports reliable hunting in unstructured environments, such as low light conditions, reflective water surfaces, and camouflage, with low latency and high success rates [45].

Drawing inspiration from the information bottleneck theory and the visual pathway of the dragonfly, we develop a neuromorphic vision hardware system with a task traction mechanism for task-adaptive rapid and robust visual perception in unstructured environments. The system consists of a 12×12 array of polarization-sensitive photodiodes composed of PVA/TAC-based linear polarizers and silicon PIN photodiodes, together with an 8×8 1T1R Ti/Pt-HfO2-Ta/Pt resistive random-access memory (RRAM) crossbar array fabricated by back-end-of-the-line (BEOL) processing directly atop the complementary metal-oxide-semiconductor (CMOS) circuitry. The former senses multidimensional light field information, including intensity, polarization, and wavelength, with a polarization extinction ratio of 9000:1, whereas the latter supports the temporal variation perception and analogue computation required by the task traction mechanism, with a wide programmable conductance range and retention time exceeding 100 ks.

The task traction mechanism consists of three sequential modules as shown in Figure 1(c): In the feature traction, light field information sensed by the polarization-sensitive photodiodes enters the nonvolatile conductance states of the RRAM to compute a gradient-based metric for polarization and light intensity images, enabling dynamic selection of the sensing information most relevant to the current scene. Subsequently, attention traction exploits transient temporal cue in the light field via dynamic conductance modulation of the RRAM to extract region of interest (ROI). Finally, prediction traction harnesses temporal cue retained in the conductance states of the RRAM to anticipate the target, thereby reducing system latency (Figure 1(d)). Collectively, these modules constitute a task traction mechanism realized through the hardware system at the perceptual front end. By preserving task-oriented information while suppressing redundant input, the neuromorphic vision system enables visual intelligence in open-world environments, with an average visual task execution time of 193 $\mu$s.

To evaluate the effectiveness of the proposed system, we conduct experiments across eight open-world scenarios, including specular reflection, water reflection, scattering, camouflage, low light, high dynamic range, transparent interference, and illumination variation. By exploiting task-relevant information, the system achieves accuracy improvements of 25.54%, 37.73%, and 36.10% for object tracking, object segmentation, and trajectory prediction, respectively, together with a 30.6-fold latency reduction relative to state-of-the-art (SOTA) machine vision baselines [16–25]. Finally, we validate the system on a vehicle-mounted platform for autonomous driving, where it outperforms existing solutions, including RGB cameras [46], infrared cameras [47], LiDAR [48], millimeter wave radar [49], and event cameras [50], establishing a feasible route toward rapid and robust perception in unstructured environments.

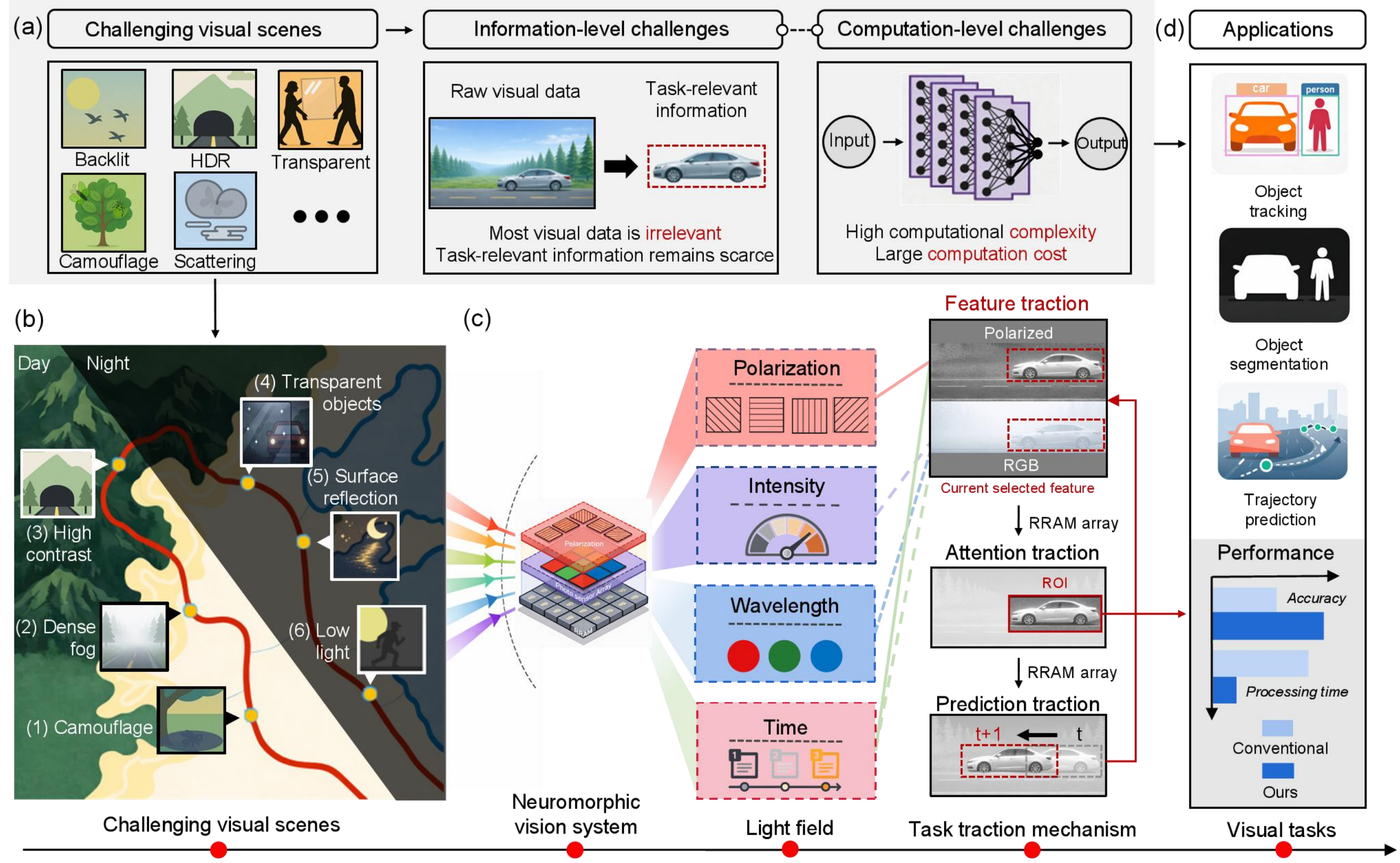

**Figure 1 The challenges of open-world unstructured visual perception and the solution with the task traction mechanism.** (a) Diverse and unpredictable visual conditions in open-world unstructured scenes, such as backlight, high dynamic range, transparent interference, camouflage and scattering, pose various challenges across both information and computation levels. (b) Representative autonomous driving scenarios under day and night conditions. (c) Task traction mechanism for in-sensor information distillation. Multidimensional light field information are progressively processed through feature traction for light field selection, attention traction for region of interest extraction, and prediction traction for forthcoming object trajectory. (d) Only distilled task-relevant information is delivered to visual tasks, enabling object tracking, object segmentation, and trajectory prediction with higher accuracy and lower processing time than whole-field global calculation.

## Results

### *System overview*

The neuromorphic vision system is composed of a polarization-sensitive imager, neuromorphic processor and FPGA (Figures 2(a)-2(c)). The polarization-sensitive imager comprises a 12×12 polarizer array and a 12×12 photodiode array, with the former decomposing incident light into four polarization components at 0°, 45°, 90°, and 135°, and the latter converting the light components into electrical signals. These signals are grouped into four-channel units and sent into the light field processing circuit, where intensity and polarization information are delivered to the computation region of the RRAM array to obtain their gradient features (i.e., feature traction) and select the light field that is relevant to the current scene. The selected information is then routed to the temporal feature extraction circuit, which extracts temporal features of the target and converts them into modulation voltages. These voltages are applied to the perception region of the RRAM array, driving conductance state changes that serve as temporal cues (i.e., attention traction), which are subsequently read out and transferred to the FPGA for ROI generation. Subsequently, the temporal cues are fed back into the computation region of the RRAM array to derive motion intensity, which is combined with the ROI to anticipate the target (i.e., prediction traction). Finally, the selected light field from the light field processing circuit and the ROI are jointly processed in the FPGA for downstream visual tasks. More detailed descriptions of the task traction mechanism and the heterogeneous information processing implemented across the computation, perception, and monitoring regions of a single RRAM array are presented in section 'Methods' and SNote 5.

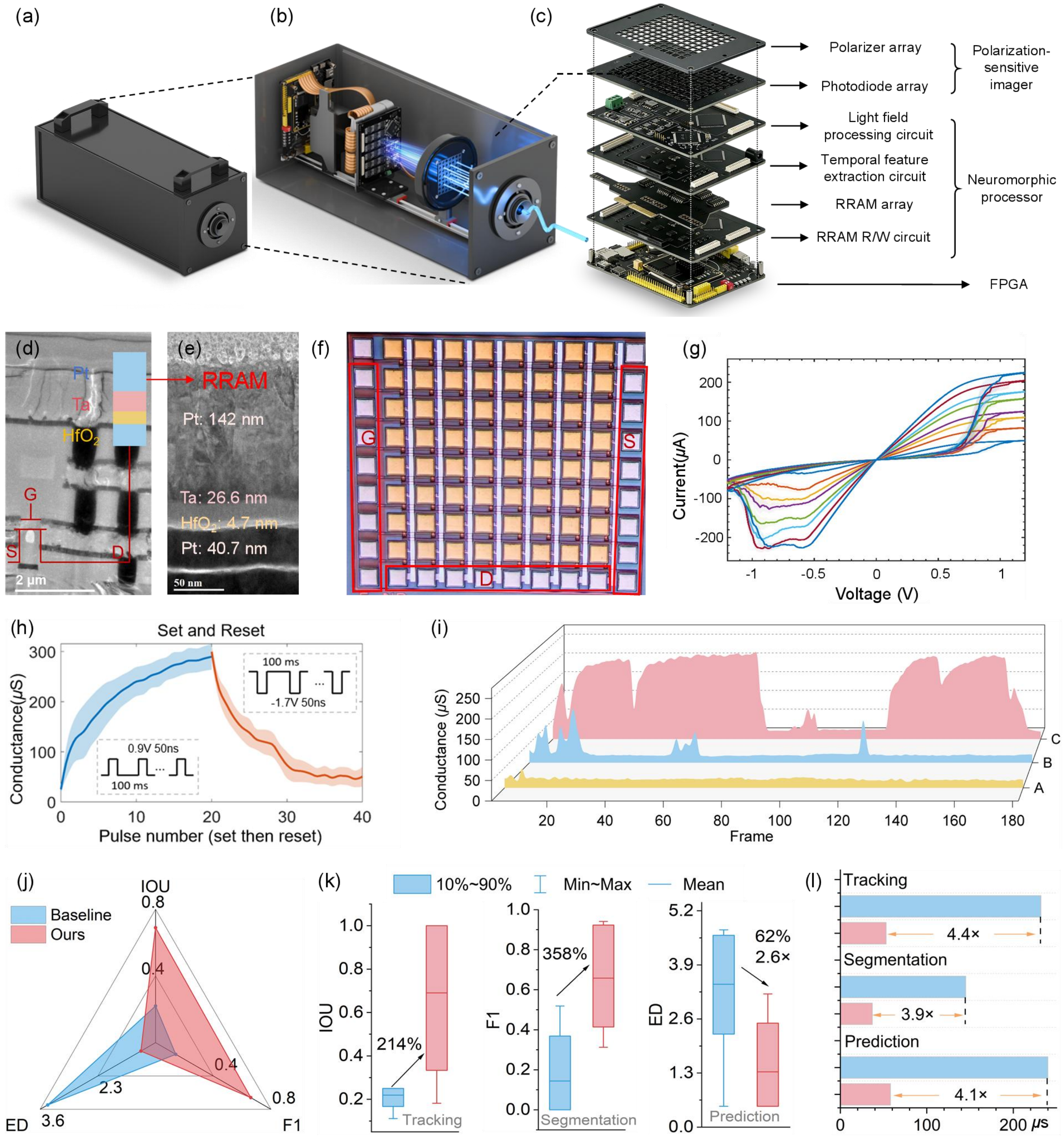

**Figure 2 Hardware implementation and proof of concept validation of the neuromorphic vision system.** (a,b) Exterior and internal views of the integrated vision system, comprising a polarization sensitive imager, neuromorphic processor and FPGA. (c) Exploded view of the hardware architecture, including the polarizer array, photodiode array, light field processing circuit, temporal feature extraction circuit, RRAM array, RRAM read and write (R/W) circuit and FPGA. (d,e) Cross sectional electron microscopy images of a representative 1T1R RRAM cell and the Ti/Pt-$HfO_2$-Ta/Pt device stack. (f) Optical micrograph of the fabricated 1T1R RRAM array. (g) Current-voltage characteristics of a representative 1T1R cell measured under different gate voltages, showing reliable resistive switching behavior. (h) Reversible analogue conductance modulation of a representative RRAM cell under successive set and reset voltage pulses. (i) Experimentally measured conductance evolution of representative RRAM cells in the integrated neuromorphic vision system during proof-of-concept validation under specular interference, where A, B and C correspond to low-, medium- and high-frequency voltage modulation, respectively. (j) Radar-chart

comparison of object tracking, object segmentation and trajectory prediction using the proposed system and the full image intensity baseline. (k) Quantitative performance comparison for tracking IoU, segmentation F1 score and trajectory prediction ED. (l) End-to-end latency comparison across the three downstream visual tasks.

***Design and characteristics of the polarization-sensitive imager***

The polarization-sensitive imager consists of a 12×12 linear polarizer polymer array and a 12×12 photodiode array. The polarizer units are arranged with analyzer orientations of 0°, 45°, 90°, and 135°. The polarizing film covers the 400-700 nm spectral range, with an extinction ratio of approximately 9000:1, and a single sheet transmittance is of 42 ± 2%. The photodiode array is composed of visible band silicon PIN photodiodes, each with a photosensitive area of 7.5 $mm^2$, a peak sensitivity at 550 nm, and a spectral bandwidth of 440-700 nm. To suppress optical crosstalk, the polarizer to photodiode spacing is set to 0.45 mm, minimizing lateral displacement under oblique incidence. Photocurrents from the four channels are converted into stable voltage outputs by transimpedance amplifier readout circuits for subsequent light field information processing. Finally, the assembled polarization sensitive photodetector is characterized using integrating spheres, yielding degree of polarization, angle of polarization and light intensity's RMSE values at 0.03 (3.00% error), 3.36° (1.87%), and 3.13%, respectively. More details are provided in Figure S1.

***Fabrication and characterization of RRAM array***

The RRAM array used in this work is fabricated through a BEOL compatible process integrated with the CMOS circuitry, as illustrated in Figure 2(d). Initially, the bottom electrode, composed of a 5 nm Ti adhesion layer and a 40 nm Pt layer, is deposited by magnetron sputtering and patterned via photolithography and lift-off. Subsequently, a 5 nm $HfO_2$ dielectric layer is deposited by atomic layer deposition (ALD). Contact vias are then opened through the dielectric layer using photolithography and inductively coupled plasma (ICP) etching. Finally, the top electrode stack, consisting of a 30 nm Ta layer and a 150 nm Pt layer, is deposited by magnetron sputtering and patterned to complete the device structure, as shown in Figure 2(e). An optical photograph of the 1T1R array is shown in Figure 2(f). Figure 2(g) presents the I-V characteristics of a representative 1T1R cell measured under eight distinct gate voltage conditions, with 20 cycles recorded for each condition, clearly revealing the SET and RESET transitions. Using a write-verify scheme, the fabricated RRAM array can be reliably programmed into multiple distinguishable analogue conductance states with a programming accuracy within 5 $\mu$S, corresponding to ~1.5% of the full conductance window, and retention exceeding 100 ks (see Figures S2-S5). The RRAM array also exhibits fast switching under 30 ns programming pulses and reliably undergoes SET and RESET transitions, see Figures S6-S7 and Figure 2(h). Figure 2(i) and Figures S8 display that visual input variations can be converted into corresponding RRAM conductance evolution through voltage modulation. Benefiting from rapid conductance modulation and long retention time, the RRAM array enables heterogeneous information processing across distinct functional regions of a single array, as detailed in the Methods section.

***Validation in unstructured open-world scenes***

After integrating the hardware components, we evaluate the neuromorphic vision system under specular interference as a proof-of-concept validation of the proposed task traction mechanism, as shown in Figure S9. Object tracking, object segmentation and trajectory prediction are compared with a full-image light-intensity baseline using intersection over union (IoU), F1 score and Euclidean distance (ED) as the corresponding metrics, respectively, with details provided in SNote 6. Compared with the baseline implemented on an NVIDIA GeForce RTX 4080 GPU, the fully hardware system improves the accuracy of tracking, segmentation and trajectory prediction by 214%, 358% and 62%, respectively. At the same time, it reduces the task latency to 193 $\mu$s, corresponding to a 4.6-fold decrease, and achieves approximately 128-fold lower power consumption. Detailed comparisons are provided in Supplementary Note 13 and Supplementary Tables 1 and 2.

To assess the scalability of the proposed neuromorphic vision system (SNote 7) in more complex open-world scenes, we establish a benchmark spanning eight challenging conditions: specular reflection, water reflection, scattering, camouflage, low light, high dynamic range, transparent interference, and complex illumination. Using this benchmark, we compare the proposed method with SOTA approaches for object tracking, object segmentation, and trajectory prediction in simulation. The baselines encompass four representative models: (1) a base model that directly processes the original visual input via Farneback [51] (SNote 8), (2) an enhancement model that applies scenario-specific image enhancement before the base model, including eight methods [16-23] such as DSRNet for reflection removal [17] and CIDNet for low light enhancement [19], (3) an end-to-end model that directly maps input images to outputs (YOLOv11 [11]), and (4) a multi-sensor fusion model that fuses different sensing modalities (Mask-DiFuser [24] and MoETrack [25]). Detailed descriptions of the benchmark scenarios and the baseline methods [11,16-25, 51] are provided in SNotes 9 and 10.

Representative visual comparisons across the eight challenging scenarios are shown in Figure 3 and Figures S10-S17. Relative to the base model (denoted as RGB for clarity), the proposed system improves average task accuracy by 724.17%, 424.47%, and 74.31% for object tracking, object segmentation, and trajectory prediction, respectively. For latency evaluation, the end-to-end processing time is reduced by 6.09×, 5.26×, and 5.55× across the three tasks, respectively, as shown in Figures S18-S20. To further assess its competitiveness against established solutions for challenging visual perception, we evaluate representative methods and denote the best-performing result among them as SOTA. Comprehensive results for all evaluated methods are provided in STables 3-10. Relative to SOTA, the proposed system achieves average accuracy gains of 25.54%, 37.73%, and 36.10% for the three tasks, respectively, while reducing latency by 32.69×, 29.54×, and 29.86×.

These results indicate that the task traction mechanism improves task performance and computational efficiency by focusing processing on task-relevant regions. The effectiveness of each traction module, including feature traction, attention traction and prediction traction, is evaluated in Figures S21-S25. In addition, ablation experiments show that individually removing the three traction modules reduces average accuracy by 37.92%, 6.84%, and 31.97%, respectively, while increasing latency by 3.08×, 5.64×, and 3.26×. Detailed quantitative analyses are provided in Figure S26 and STable 11. In addition, we also evaluate the performance of the task traction mechanism under RRAM array variability, and the results show that it is insensitive to moderate device variations, as shown in Figure S27.

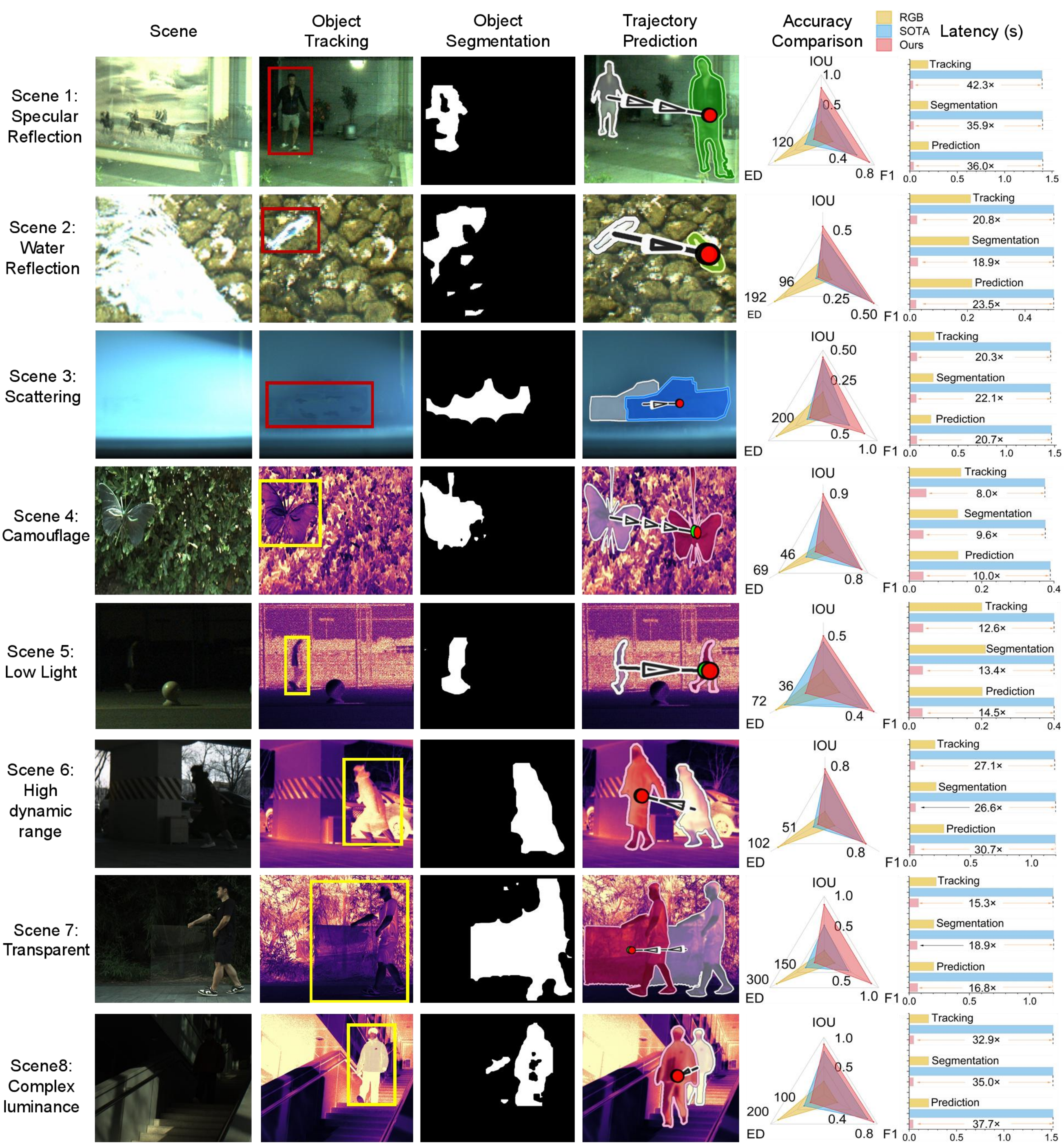

**Figure 3 Performance of the neuromorphic vision system across diverse open-world unstructured scenes.** Visualization of three downstream tasks across eight unstructured environments, including specular reflection, water reflection, scattering, camouflage, low light, high dynamic range, transparent interference, and illumination variation based on simulated RRAM array. From left to right, representative input scenes, object tracking, object segmentation, trajectory prediction, accuracy comparisons against base model and SOTA (the best-performing result among all evaluated methods), and latency comparisons for the three visual tasks.

***Validation in autonomous driving***

To further evaluate the applicability of the neuromorphic vision system in practical scenarios, we mount the hardware system on the roof of a vehicle and examine its performance in autonomous driving under high dynamic range conditions, as shown in Figure S28. Compared with the baseline method using full-image light intensity information, the system improves object tracking, object segmentation, and trajectory prediction by 329%, 1116%, and 63%, respectively, while reducing end-to-end latency by 4.1-fold, 4.6-fold, and 3.9-fold. More detailed results are provided in Figures S29-S31, and STable 12.

To assess the scalability of the proposed neuromorphic vision system in more complex autonomous driving scenarios, we further conduct evaluations in an underground parking garage characterized by low illumination and pronounced luminance variations, which substantially degrade visual quality and make moving vehicles and pedestrians difficult to identify using conventional perception systems, as shown in Figures 4(a)-(c) [52, 53]. The qualitative results in Figures 4(d)-4(f) show that the neuromorphic vision system enables reliable perception performance. Quantitative comparisons with SOTA under the same experimental condition are summarized in Figure 4(g). As detailed in Figure 4(h), object tracking accuracy boosts approximately 195%, object segmentation performance increases by 121%, and trajectory prediction error decreases by 80%. In addition to accuracy gains, the task traction mechanism significantly reduces computational latency. As shown in Figure 4(i), the end-to-end processing time for tracking, segmentation, and prediction decreases by 17×, 18×, and 17×, respectively. More detailed results are provided in STable 13. These results demonstrate that the task traction mechanism supports reliable and time-efficient visual perception in challenging autonomous driving scenes.

We further compare the proposed approach with representative sensing modalities used in autonomous driving, including RGB cameras [46], infrared cameras [47], LiDAR [48], millimeter-wave radar [49], and event cameras [50], as summarized in Figure 4(j). Our work demonstrates robust performance across nine conditions, including transparent obstacles, camouflage and complex illumination. In contrast, existing sensing modalities exhibit limitations. For instance, RGB cameras are highly sensitive to illumination variation, leading to significant degradation under low light or high dynamic range conditions [18]. LiDAR and millimeter-wave radar infer object positions primarily from reflected signals, therefore in enclosed environments, multi-path reflections from walls and structural surfaces introduce ghost measurements and localization ambiguity [54, 55]. In thermally homogeneous underground environments, infrared cameras provide limited discriminative cues owing to their reliance on thermal contrast [56]. Event cameras are effective for detecting fast moving targets, whereas static or slowly moving objects may generate insufficient event responses, potentially compromising autonomous driving safety, as in the case of stationary pedestrians or stopped vehicles [57].

These results and comparisons demonstrate that the proposed neuromorphic vision system enables robust, accurate and low-latency perception in challenging autonomous driving scenarios. This advantage arises from the task traction mechanism, which preserves task-relevant information while compressing the input, providing a promising visual foundation for autonomous systems.

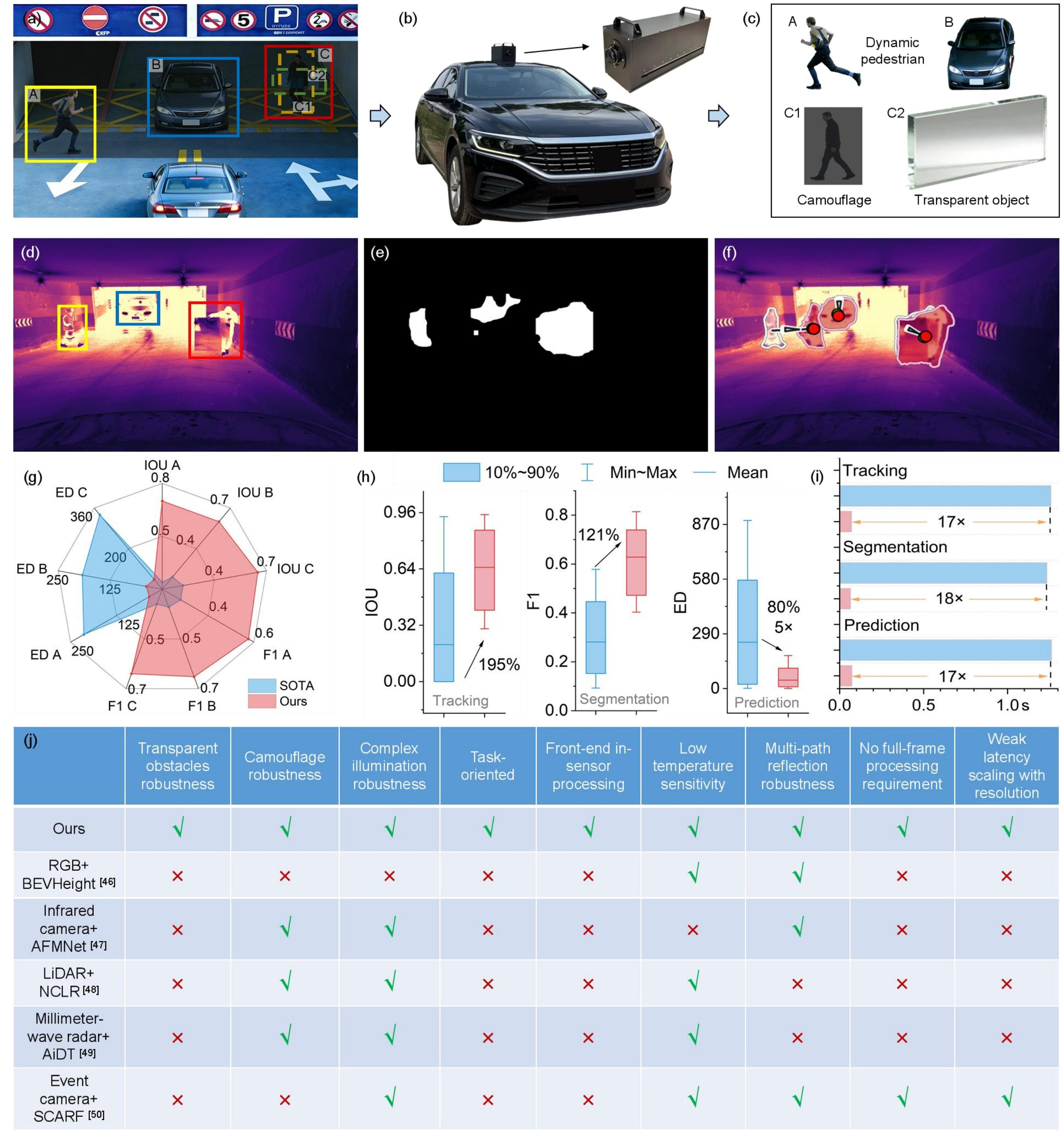

| | Transparent obstacles robustness | Camouflage robustness | Complex illumination robustness | Task-oriented | Front-end in-sensor processing | Low temperature sensitivity | Multi-path reflection robustness | No full-frame processing requirement | Weak latency scaling with resolution |
|---|---|---|---|---|---|---|---|---|---|
| Ours | √ | √ | √ | √ | √ | √ | √ | √ | √ |
| RGB+ BEVHeight [46] | × | × | × | × | × | √ | √ | × | × |
| Infrared camera+ AFMNet [47] | × | √ | √ | × | × | × | √ | × | × |
| LiDAR+ NCLR [48] | × | √ | √ | × | × | √ | × | × | × |
| Millimeter-wave radar+ AiDT [49] | × | √ | √ | × | × | √ | × | × | × |
| Event camera+ SCARF [50] | × | × | √ | × | × | √ | √ | √ | √ |

**Figure 4 Autonomous driving evaluation of the neuromorphic vision system against SOTA perception methods and sensing modalities.** (a–c) Representative autonomous driving scenes in an underground parking garage under low illumination and pronounced luminance variation, where pedestrians, vehicles and transparent obstacles are difficult to perceive using conventional vision systems. (d–f) Qualitative result of the proposed work applied to three downstream visual tasks based on simulated RRAM array. (g) Quantitative performance comparison with SOTA methods (the best performing result selected from four perception strategies) in terms of tracking IoU, segmentation F1 score and trajectory prediction ED. Objects A, B, and C are defined as shown (a).(h) Statistical comparison of overall performance for tracking, segmentation and trajectory prediction. (i) End-to-end latency comparison for the three downstream visual tasks. (j) Comparison with autonomous driving sensing modalities across nine conditions, including transparent obstacles, camouflage and complex illumination.

## Discussion

***Comparison with existing computer vision and artificial vision solutions***

Current perception approaches for unstructured open-world environments can be generally categorized into two routes. The first route improves perception reliability by increasing data scale or model complexity to adapt to diverse scene conditions [11-25]. Ithaca365 improves autonomous driving perception under challenging conditions by scaling up the training set, yielding a dataset of 682K images [13]. FlowIE [18] and CIDNet [19] improve degraded inputs by introducing image restoration models before downstream perception, whereas SAM 2 [12] improves performance by scaling up model capacity to 224.4M parameters. However, their performance degrades sharply when scenes differ substantially from the training dataset. Moreover, larger models significantly increase end-to-end latency in the perception pipeline (SAM2 requires 390 ms to segment a 960 × 540 image on a Tesla T4 GPU).

The second route incorporates advanced visual sensing, enabled by photosensitive materials and devices, to provide additional light field information beyond conventional light intensity imaging [26-35]. For example, geometric filterless photodetectors have been designed to provide polarization sensitive infrared cues that benefit target discrimination and material identification under weak intensity contrast [30]. Hyperspectral image sensors integrate wavelength selective material and filter structures to improve object recognition, material segmentation and scene understanding [33,34]. However, their performance are tailored towards specific conditions and remains limited in more general, open-world visual tasks.

Our results demonstrate that the proposed task traction mechanism addresses this gap by adapting perception to task demands. By selecting task-relevant light fields, extracting task regions and anticipating target trajectory, the system restricts processing to task-relevant cues, thereby mitigating performance degradation caused by task-irrelevant inputs in unstructured environments and enabling reliable, time-efficient open-world perception. Other light field modalities beyond polarization and intensity, including infrared and spectral information are also compatible within the framework introduced in this work.

***Comparison with existing neuromorphic approaches for vision***

Recently, a wealth of research has leveraged the non-volatile state retention of neuromorphic devices (such as phase-change memory and RRAM) to accelerate computation in the context of machine vision [58-62]. In situ spectral reconstruction is implemented on a memristor chip by physically mapping the spectral decoding process into memristor in [59], achieving a 26.5× speedup. Spike-based dynamic computing is realized on an asynchronous sensing and computing neuromorphic chip, where sparse event-driven signals are processed with spiking neural networks for efficient computation [63]. However, these approaches focus on post-perception inference and computation, whereas the inference in vision tasks can involve a large number of parameters for whole-image processing, which can contribute to considerable computational cost and latency. By contrast, the proposed system uses neuromorphic hardware to preserves task-relevant information and suppresses redundant background responses at the perception front end, thereby reducing the computational burden of downstream visual tasks.

Furthermore, to implement this mechanism at the sensor front end, we design the RRAM array to support selective perception and in-sensor computation within a single array. This design requires the RRAM array cells to support both rapid conductance updates (for spatiotemporal variation perception) and stable conductance retention (analogue gradient computation). This capability can be attributed to the coordinated design of the material stack, switching interfaces, and 1T1R circuit architecture. Specifically, the 5 nm $HfO_2$ switching layer deposited by atomic layer deposition provides a highly uniform and dense dielectric matrix that minimizes defect clustering and ensures consistent switching behavior across the array. This ultrathin dimension enables strong electric field confinement at low operating voltages, which accelerates oxygen ion drift and supports 30 ns conductance modulation. The reactive Ta layer at the top electrode interface functions as an oxygen reservoir that promotes a localized oxygen-vacancy-rich region near the Ta/$HfO_2$ interface. This interfacial vacancy distribution reduces the energy barrier for conductive filament formation and rupture, contributing to the observed high-frequency switching capability. The chemically inert Pt bottom electrode provides a stable nucleation surface that helps localize filament growth and suppresses uncontrolled interfacial reactions, supporting long-term retention and endurance. The relatively thick top electrode stack (30 nm Ta and 150 nm Pt) ensures low series resistance, which helps minimize parasitic RC delays during high-frequency operation. Integration within a 1T1R architecture provides precise compliance current control through the access transistor, preventing excessive filament growth and enabling accurate, incremental conductance updates required for analogue neuromorphic temporal coding. Together, these design choices in materials selection, interface engineering, and circuit integration collectively enable the RRAM array to achieve both rapid dynamic response and stable conductance retention.

***General principles for task-oriented perception***

More fundamentally, our results suggest two basic principles that extend beyond vision to general perceptual processing. First, task demands should guide information acquisition at the perceptual front end. In complex environments, task-irrelevant signals increase computational cost and interfere with task-relevant information. A perceptual system should therefore prioritize task-relevant information at the front end, while suppressing interfering signals before they propagate into downstream computation. Second, prediction should guide subsequent perception. By anticipating targets trajectories, sensing can focus on task-relevant regions, thereby enriching useful information and reducing response time, as exemplified by dragonfly prey interception [43]. Together, these principles outline a general strategy for efficient, task-oriented perception.

For example, in auditory perception within a crowded urban environment, multiple overlapping sounds such as conversations, traffic, sirens, and construction noise are simultaneously present. The task can prioritize certain sound components from the outset, such as the voice of a specific speaker, the horn of an approaching vehicle, or an emergency alarm, rather than processing all acoustic inputs equally. By focusing on task-relevant signals first, the system reduces interference from irrelevant background noise and improves recognition accuracy and processing efficiency. Prediction can then guide the next sensing step toward the expected spatial location, temporal window, or frequency range of the relevant source, based on recent changes in the signal, so that sensing and computation concentrate on the most informative channels. This approach shortens response time and stabilizes perception, even when the auditory scene changes rapidly and unpredictably. A similar principle can be applied to tactile perception during object manipulation. Instead of processing all contact signals equally, the task can prioritize tactile inputs that are most informative for grasp stability, such as sudden slip, rapid changes in pressure, or edge contact near the fingertips. Prediction can then be used proactively to direct subsequent sensing toward regions likely to experience instability, thereby enabling anticipation of incipient slip or object displacement and corrective adjustment of the grip before full grasp failure occurs. Taken together, these examples suggest that these two principles may extend beyond vision and provide a general framework for efficient perceptual processing.

## Method

### *Task traction mechanism*

Enabled by the programmable, non-volatile, and rapidly modulated conductance states of RRAM, the task traction mechanism is implemented within a single functionally partitioned RRAM array to distill task-relevant information in unstructured scenes, as illustrated in Figures 5(a) and 5(b). The computation region utilizes programmable and retained conductance states to support stable analogue operations, including gradient-based feature evaluation for feature traction and motion-intensity estimation for prediction traction. The perception region exploits rapid conductance modulation to encode temporal variations of the selected light field into RRAM states, thereby generating temporal cues for attention traction. In parallel, the monitoring region provides a lightweight pathway for detecting unexpected motion outside the predicted task region. This configuration within a single RRAM array enables adaptive feature selection, temporal cue formation, and trajectory prediction to be integrated at the sensor front end, as schematically shown in Figures 5(c)-5(f).

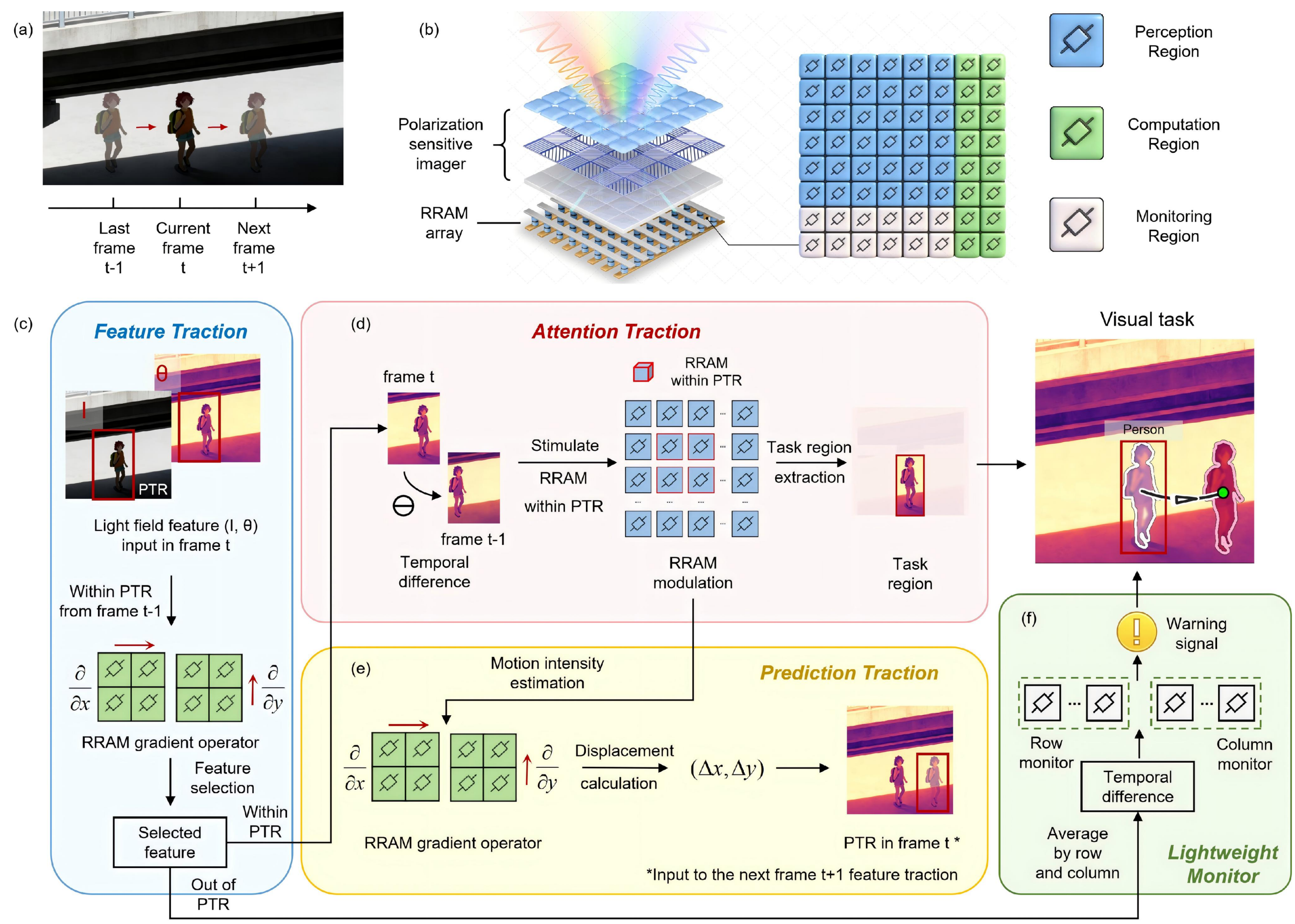

**Figure 5 Task traction mechanism of the neuromorphic vision system for task-relevant information distillation.** (a) In unstructured scenes, task-relevant targets and informative visual cues vary over time and cannot be specified a priori, resulting in substantial task irrelevant information redundancy and inefficient computation. (b) Schematic of the neuromorphic vision system and functional partitioning of the RRAM array into perception, computation and monitoring regions. (c) Feature traction: task-relevant light field is selected within a prediction task region (PTR) based on gradient features obtained in the computation region of the RRAM array. (d) Attention traction: temporal features of the selected light field modulate the perception of RRAM array to extract the task region (ROI). (e) Prediction traction: motion intensity is derived from temporal cues through computation region in the RRAM array to estimate target trajectory and update the prediction task region for the next frame. (f) Lightweight monitoring: non-task regions are monitored in the monitoring region of the RRAM array to detect unexpected motion and determine whether additional ROIs should be introduced for multi-target perception.

### *Feature Traction*

In the feature traction, we design an adaptive feature selection (polarization and intensity) strategy based on gradient entropy(see Figure 5(c), and the physical interpretation and calculation process of the light field information are provided in SNote 11 and 12). Specifically, the light field information are firstly extracted from the prediction task region (PTR) generated by the prediction traction in the previous frame to obtain the local information within prediction task region:

$$F_t^{local}(x,y) = M_{t-1}^{p}(x,y) \otimes F_t(x,y) \tag{1}$$

$F_t$ is the sensing light field information, contain polarization ( $\theta$ ) and intensity ( $I$ ), and $M_{t-1}^{p}(x,y)$ is the prediction task region from last frame.

Subsequently, the gradient of the light field information within the prediction task region is computed. Specifically, prior to task execution, the RRAM array is programmed with voltage pulses to predefined conductance states, allowing it to operate as a stable gradient operator based on its non-volatile property. In this work, the gradient is computed using a Roberts Cross operator (as shown in Figure 6(a)):

$$C_x = \begin{bmatrix} 1 & 0 \\ 0 & -1 \end{bmatrix},\ C_y = \begin{bmatrix} 0 & 1 \\ -1 & 0 \end{bmatrix} \tag{2}$$

It should be noted that other gradient operators can also be adopted within the same framework. For gradient kernels containing negative weights, the operator is decomposed into two non-negative kernels, and the signed gradient response is obtained through their subtraction:

$$C_x = C_x^+ - C_x^- = \begin{bmatrix} 1 & 0 \\ 0 & 0 \end{bmatrix} - \begin{bmatrix} 0 & 0 \\ 0 & 1 \end{bmatrix}, C_y = C_x^+ - C_x^- = \begin{bmatrix} 0 & 1 \\ 0 & 0 \end{bmatrix} - \begin{bmatrix} 0 & 0 \\ 1 & 0 \end{bmatrix} \tag{3}$$

Therefore, the gradients in the $x$ and $y$ directions can be calculated as follows (as shown in Figure 6(a)):

$$G_x = C_x^+ \otimes F_t^{local} - C_x^- \otimes F_t^{local}, G_y = C_y^+ \otimes F_t^{local} - C_y^- \otimes F_t^{local} \tag{4}$$

After discretization and Shannon entropy calculation, the gradient entropy is obtained. Finally, feature traction selects the light field information with the maximum gradient entropy at local $P_t^{local}$ and global scales $P_t^{global}$. The former is used for subsequent traction, whereas the latter is retained for monitoring.

***Attention Traction***

In the attention traction, a high-pass filtering circuit extracts the temporal variation of the selected local feature. The response magnitude is then converted by different operational amplifier into positive or negative modulation pulses, which are used as the modulation voltage applied to the RRAM array. This process can be expressed as:

$$V_{i,j}^{\mathrm{mod}} = g\left|h\left(P_t^{local}(i,j), P_{t-1}^{local}(i,j)\right)\right| \tag{5}$$

Where $h(\cdot)$ is the high-pass filtering operation, and $g(\cdot)$ denotes the piece-wise stimulation function used to switching trends in the RRAM states, thereby enhancing task-relevant dynamic regions while suppressing background fluctuations and noise. It is important to note that for high-resolution images, such as 1224×1024, the image is divided into multiple $m \times n$ units, where $m = 204$ and $n = 170$ in this work. The signal derived from each unit is then used to modulate the RRAM conductance. For example, when a pedestrian moves in front of the vehicle (Figures 6(b) and 6(c)), a pronounced voltage variation is observed at position (3, 4), generating a pulse (Figures 6(d) and 6(e)) that drives the RRAM device into a high conductance state (Figure 6(f)), and the RRAM state is denoted as $S_t$, serving as the temporal cue. The RRAM state $S_t$ is then converted into the task regions in two steps. First, binarization is performed using a predefined threshold. Second, morphological opening and closing are applied to refine the binarized regions, and connected regions with an area larger than a predefined pixel number threshold are retained as ROIs. This process produces one or multiple task regions, as shown in Figure 6(g). The task region map $M_t$ is defined as a binary map, where task-relevant regions are labeled as 1 and non-relevant regions as 0. Therefore, the current task-relevant features $P_t^R$ can be expressed as:

$$P_t^R(x,y) = M_t(x,y) \otimes P_t^{local}(x,y) \tag{6}$$

Subsequently, $P_t^R$ is fed into downstream tasks, such as object tracking, object segmentation, and trajectory prediction.

***Prediction Traction***

The prediction traction leverages the task region and the RRAM conductance state in the current frame to predict the task region in the subsequent frame, as shown in Figure 5(e).

At time $t$, the horizontal and vertical variations of the RRAM state within the task region (denoted as $S_t^m$) reflect the motion intensity of the target. These variations are computed using an RRAM-based gradient operator, as illustrated in Figure 6(a), and expressed as:

$$G_x^{motion} = \frac{\partial S_t^m(x,y)}{\partial x}, G_y^{motion} = \frac{\partial S_t^m(x,y)}{\partial y} \tag{7}$$

The average gradient within $M_t$ is used to quantify the motion intensity:

$$Q_x = \frac{1}{|M_t|}\sum \left|G_x^{motion}(x,y)\right|, Q_y = \frac{1}{|M_t|}\sum \left\|G_y^{motion}(x,y)\right\| \tag{8}$$

Where $|M_t|$ denotes the number of pixels within the task region. $Q_x$ and $Q_y$ are regarded as the motion intensities of the target along the horizontal and vertical directions, respectively, and are mapped to the spatial displacement of the target's center position through a predefined transformation:

$$\Delta x = f(Q_x), \Delta y = f(Q_y) \tag{9}$$

Where $f(\cdot)$ is the mapping function. Accordingly, the predicted center position of the target at the next time step is:

$$p_{t+1} = (x_t + \Delta x_t, x_y + \Delta x_y) \tag{10}$$

Where $(x_t, y_t)$ represents the center position of $M_t$ at the current frame $t$. Therefore, preliminary prediction task region in frame $t$ is generated as $\tilde{M}_t^p$, centered at $p_{t+1}$ with the same height and width as $M_t$. Subsequently, the final prediction task region $M_t^p$ is obtained by taking the union of $\tilde{M}_t^p$ and $M_t$ as:

$$M_t^p = \tilde{M}_t^p \cup M_t \tag{11}$$

***Lightweight non-task region monitoring***

To monitor unexpected motion in non-task regions and determine whether additional ROIs should be added, we design a lightweight surveillance pathway. The overall mechanism consists of two stages.

In the first step, the non-task region is partitioned into a $p \times q$ grid to obtain a coarse-grained representation for the detection of unexpected motion, in this work $p = q = 6$. Subsequently, row-wise and column-wise sums are computed to form low-dimensional monitoring indicators:

$$\begin{aligned} \overline{r}_i &= \sum_{j=1}^{q}(1 - M_{i,j}^t)(P_t^{global})_{i,j}, \quad i = 1,2,...,p \\ \overline{c}_j &= \sum_{i=1}^{p}(1 - M_{i,j}^t)(P_t^{global})_{i,j}, \quad j = 1,2,...,q \end{aligned} \tag{12}$$

The monitoring vector can be represented as:

$$m = [\overline{r}_1, ...\overline{r}_p, \overline{c}_1, ..., \overline{c}_q] \in R^{p+q} \tag{13}$$

In the second step, the monitoring signal is generated from the temporal difference of $m$ between consecutive frames:

$$V_i^m(t) = h\left(m_i(t), m_i(t-1)\right) \tag{14}$$

Where $i$ represents the index in monitoring vector. Subsequently, the RRAM resistance state is modulated based on $V_i^m$. When $V_i^m$ is below a predefined threshold, the modulation voltage is set to $V_{recover}$, keeping the RRAM in a high resistance state and suppressing weak transient fluctuations. In contrast, when $V_i^m$ is sufficiently large and sustained, the RRAM state gradually evolves toward a low-resistance state, thereby amplifying the monitoring output. Therefore, the modulation voltage of i-th RRAM $V_i^g$ can be expressed as:

$$V_i^g(t) = \begin{cases} V_i^m(t), V_i^m(t) \geq V_{th} \\ V_{recover}, V_i^m(t) < V_{th} \end{cases} \tag{15}$$

Under the modulation of $V_i^g$, the resistance state of the RRAM changes, and the output spike voltage $V_i^{spike}$ can be expressed as:

$$V_i^{spike}(t) = \frac{R_f}{R_{RRAM}(t)} m_i(t) \tag{16}$$

Where $R_{RRAM}$ is the resistance of the RRAM, $R_f$ is a fixed resistor used to adjust the output range of $V_i^{spike}$. Figure 6(h) shows that environmental disturbances in non-task regions produce small and irregular inter-frame differential voltage fluctuations. After modulation, these weak fluctuations are suppressed, whereas sustained unexpected motion generates a pronounced output spike, as shown in Figure 6(i). When $V_i^{spike}$ exceeds a predefined threshold, a warning signal is generated,and then, in the following frame, the selected global feature is introduced into attention traction, enabling unexpected motion in non-task regions to generate additional ROIs.

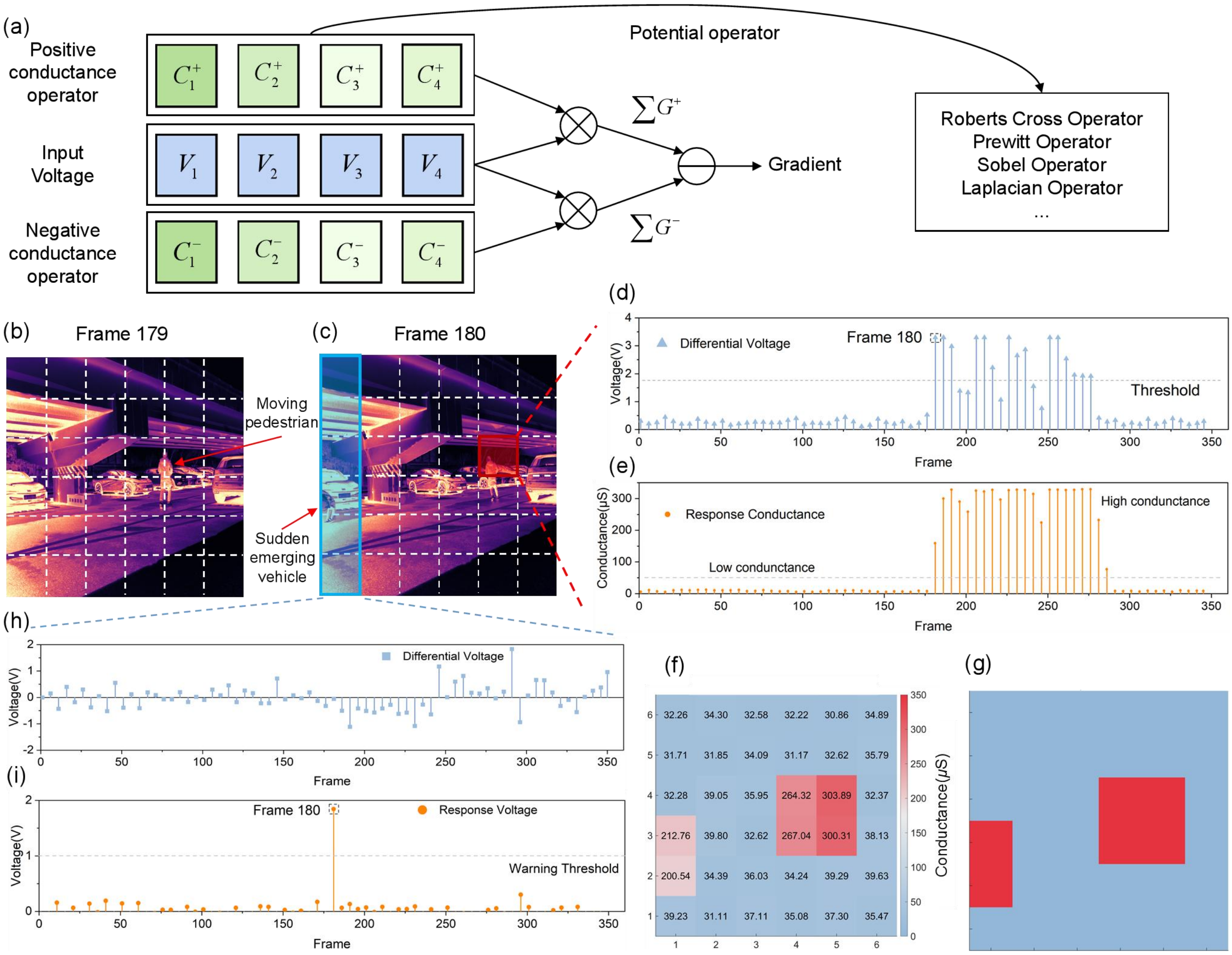

**Figure 6 Hardware operation of the task traction mechanism and lightweight monitoring pathway.** (a) RRAM-based gradient operator implemented by differential conductance states for gradient feature calculation in feature traction and motion intensity estimation in prediction traction. (b,c) Emergence of a pedestrian and the resulting localized temporal response in the corresponding spatial unit. (d) Inter-frame differential voltage, with prominent responses exceeding the threshold at frame 180. (e) Corresponding conductance evolution of the RRAM under voltage stimulation, showing the transition from low- to high-conductance states. (f) Conductance map (temporal cue) of the array after modulation. (g) Task region (i.e., ROI) obtained by binarizing the temporal cue, with the red regions denoting the extracted ROIs. (h) Differential voltage in non-task regions. (i) Output response of the lightweight monitoring pathway, which suppresses background fluctuations and generates a warning signal when unexpected motion occurs.

***Visual processing***

To demonstrate the scalability of the proposed system, visual inputs from unstructured and complex scenes are processed using simulated RRAM arrays. The switching dynamics of the RRAM are reproduced using a VTEAM model [64] calibrated with experimentally measured pulse response data (Figure S32).

***Running environment***

Performance evaluations of image enhancement methods and large-scale neural network models are conducted on an NVIDIA GeForce RTX 4080 Laptop GPU with 12 GB memory. In the proposed neuromorphic vision system, a Zynq-7020 FPGA is used to read out RRAM conductance states, generate modulation voltage and manage read/write operations.

***Consent for use of identifiable images***

All figures containing potentially identifiable images of individuals (including Figure 3,4 and 6 in main text and SNote12, Figures S10, S14-S17, S27-S28 in supplementary information) were captured with the participants' full awareness and cooperation. Written informed consent was obtained from all individuals involved, granting permission to use and publish their images in this research.